\documentstyle[prl,aps,preprint,psfig]{revtex}

\newcommand{\GeV}{\ifmmode\,{\rm GeV}\else $\,\rm GeV$\fi}
\newcommand{\cm}{\ifmmode\,{\rm cm}\else $\,\rm cm$\fi}

\begin{document}
\draft
\preprint{TU-590}
\title{Gravitino Dark Matter without R-parity}
\author{Fumihiro Takayama\footnote{e-mail: takayama@hep.phys.tohoku.ac.jp}
 and Masahiro Yamaguchi\footnote{e-mail: yama@hep.phys.tohoku.ac.jp}}
\address{Department of Physics, Tohoku University \\
         Sendai 980-8578, Japan}
\date{May 2000}

\maketitle
\begin{abstract}
  Cosmological issues are examined when  gravitino is the lightest
  superparticle (LSP) and R-parity is broken.  Decays of the next
  lightest superparticles occur rapidly via R-parity violating
  interaction, and thus they do not upset the big-bang
  nucleosynthesis, unlike the R-parity conserving case.  The gravitino
  LSP becomes unstable, but its lifetime is typically much longer than
  the age of the Universe. It turns out that observations of 
  diffuse photon background coming from radiative decays of the
  gravitino do not severely constrain the gravitino abundance, and
  thus the gravitino weighing less than around 1 GeV can be dark
  matter of the Universe when bilinear R-parity violation generates
  a neutrino mass which accounts for the atmospheric neutrino
  anomaly.
\end{abstract}

\clearpage

When one considers supersymmetric extension of the Standard Model 
\cite{Nilles}, one
often assumes R-parity conservation. The R-parity \cite{FarrarFayet} 
is a $Z_2$ parity
which distinguishes superparticles from ordinary particles. It is 
motivated to prevent too fast proton decay mediated by
dimension four operators \cite{protondecay}, but it
additionally results in a very interesting consequence in
cosmology. Namely, thanks to the R-parity conservation, the lightest
superparticle (LSP) which carries odd R-parity is stable and thus can
be considered a candidate for dark matter of the 
Universe \cite{LSPdarkmatter}. In fact
following the standard thermal history of the Universe, the lightest
neutralino (a combination of neutral gauginos and higgsinos) which
is often assumed to be the LSP tends to have abundance comparable to
the total mass density of the Universe for reasonable choices of
supersymmetry breaking parameters \cite{DMreview}.

The R-parity conservation is, however, not the only way to forbid the
dangerous proton decays caused by the dimension four operators. One can
consider R-parity violation \cite{HallSuzuki} which violates,
for example, only the lepton number as well.  In fact one of the
motivations for the R-parity violation is neutrino masses and mixing
which are strongly indicated by the atmospheric neutrino anomaly \cite{atm}
and the solar neutrino problem \cite{solar}.  When the R-parity breaks via
lepton number non-conserving interaction, a novel mechanism
\cite{HallSuzuki} of generating the neutrino masses and mixing takes place,
which has been extensively studied in the literature (for recent works, see 
\cite{r-violating,Bilinear}). Note that
R-parity itself cannot forbid dimension five proton decay operators
which might be dangerous when compared with severe experimental
bounds. There may be an alternative symmetry to the R-parity,
which manages to sufficiently suppress the nucleon decay \cite{IbanezRoss}.

When the R-parity is broken, the LSP is no longer stable, but 
decays to ordinary particles.  When the LSP is the superpartner of a
standard model particle such as  a neutralino or a slepton, the
decay of the LSP takes place very rapidly, whose typical lifetime is
much shorter than 1 sec unless the R-parity violation is extremely
small. Then the LSP cannot be the dark matter of the Universe. In
this case the gravitino decay occurs through usual supercurrent
interaction, and thus its lifetime is rather long so that it may decay during
or after the big-bang nucleosynthesis, leaving the  conventional 
gravitino problem unchanged \cite{gravitino-problem}.

In this paper, we would like to consider a different case where the
gravitino is the LSP under the assumption of the R-parity violation.
The gravitino LSP can be realized in various scenarios of
supersymmetry breaking mediation, including gauge mediation
\cite{gauge-mediation} and scenarios of low fundamental energy scale
\cite{low-energy}. Even within the conventional gravity mediation, the
gravitino can be the lightest among the superparticles for certain
choices of supersymmetry breaking parameters.

A characteristics of this case is that the lifetime of the gravitino is
very long, typically much longer than the age of the
Universe, as we will see later.
Therefore the gravitino LSP may be considered to be a dark
matter of the Universe if its decay will not cause any problems. In
fact if the decay of the gravitino LSP contains photon, which is often
the case, there comes a nontrivial constraint from the diffuse photon
background observation. We will show that the constraint is not very 
severe, allowing the gravitino to be a dark matter candidate.

To be specific, we will mainly consider the case where R-parity is
violated in bilinear terms in superpotential as well as in soft
supersymmetry breaking terms \cite{HallSuzuki,Bilinear}. 
In this case the superpotential contains
the following mass terms
\begin{equation}
W=\mu H_1 H_2 +\sum_{i=1}^{3} \mu_i L_i H_2,
\end{equation}
where $H_1$, $H_2$ are two Higgs doublets, $L_i$ ($i=1, 2, 3$) are
$SU(2)_L$ doublet leptons, and $\mu$ is a higgsino mass parameter and $\mu_i$
are R-parity violating higgsino-lepton mixing masses.  The soft SUSY 
breaking terms in the scalar potential are taken to
be of the form
\begin{eqnarray}
  V^{\mathrm{soft}}=B H_1 H_2+B_i \tilde{L}_i H_2
 +m_{H_1}^2H_1H_1^{\dag}+m_{H_2}^2H_2H_2^{\dag}
  +m_{HL_i}^2\tilde{L}_iH_1^{\dag}+m_{L_{ij}}^2
  \tilde{L}_i\tilde{L}_j^{\dag} +\cdots,
\end{eqnarray}
where we have written only bilinear terms explicitly (in a self explanatory
notation). Here $B_i$ and
$m_{HL_i}^2$ break the R-parity. The R-parity violating terms induce
non-vanishing vacuum expectation values for sneutrinos, and the
resulting neutrino-neutralino mixing yields a non-zero neutrino mass due
to a weak-scale seesaw mechanism. Neutrino masses
are also generated by loop diagrams as well, and as a consequence one can
explain both the atmospheric neutrino anomaly and the solar neutrino problem
in this framework. When the supersymmetry breaking scale is around 
100 GeV, the ratio $\sqrt{\mu_1^2+\mu_2^2+\mu_3^2}/\mu$ should be $\sim
10^{-4}-10^{-6}$ to give the neutrino mass which explains the atmospheric
neutrino anomaly \cite{Bilinear}.

Before discussing  decays of the  gravitino LSP, we would like to briefly
mention  decays of the next lightest superparticle 
(NLSP) which is now the lightest superpartner
among the Standard Model particles. In the R-parity conserving case,
the NLSP decays into gravitino with some electromagnetic and/or
hadronic activities. The decay width of the NLSP is roughly of the order
\begin{equation}
  \Gamma_{\rm{NLSP}} \sim
   \frac{1}{ 16 \pi} \frac{m_{\rm{NLSP}}^5}{M_{pl}^2 m_{3/2}^2}
\end{equation}
assuming that the decay is a two-body decay. Here 
$M_{pl} \simeq 2.4 \times 10^{18}$ GeV is the
reduced Planck scale, 
$m_{3/2}$ is the gravitino mass, and $m_{\rm{NLSP}}$ is the NLSP mass. 
If the decay
occurs during or after the big-bang nucleosynthesis epoch, which is
the case for a relatively heavy gravitino,  the
abundances of the light elements will be significantly changed.  This
issue was discussed in Refs.\cite{MMY,GGR,AHS} and an upper bound on the
gravitino mass is obtained, provided that the Universe follows the standard
thermal evolution.

Now if we consider the R-parity violation, the situation
changes drastically.  The NLSP decays into ordinary particles
({\it i.e.} R-parity even particles) via the R-parity violating
interaction,  and its lifetime becomes much shorter than 1 sec. 
Thus the decay of the NLSP becomes harmless \cite{GGR}.

Let us next turn to the decay of the gravitino LSP. The lifetime of
the gravitino is long because it experiences the gravitational
interaction suppressed by $M_{pl}$ and also the small R-parity
violating coupling is involved for the LSP decay. In fact the lifetime
is typically much longer than the age of the Universe. To see this,
let us consider the bilinear R-parity violation with the heaviest
neutrino mass fixed around 0.07 eV. We assume that the lightest
neutralino is bino-dominant. Then the dominant decay mode of the
gravitino LSP is into a photon and a neutrino.  This decay occurs
through the interaction
\begin{equation}
    L_{int}=-\frac{i}{8M_{pl}}\bar{\psi}_{\mu}
          [\gamma^{\nu}, \gamma^{\rho}] \gamma^{\mu} \lambda F_{\nu \rho},
\end{equation}
where $\psi_{\mu}$ is the gravitino field, $F_{\nu \rho}$ is the
field strength for the photon, and $\lambda$ represents the superpartner of
the photon, ``photino'', which  contains a neutrino component via
neutralino-neutrino mixing after the  sneutrino develops the vacuum
expectation value.  Thus we  evaluate the lifetime of the gravitino as follows:
\begin{eqnarray}
  \Gamma(\tilde{G}\to \gamma \nu)=
  \frac{1}{32\pi}|U_{\gamma\nu}|^2
  \frac{m_{3/2}^3}{M_{pl}^2}\left(1-\frac{m_{\nu}^2}{m_{3/2}^2}\right)^3
  \left(1+\frac{1}{3}\frac{m_{\nu}^2}{m_{3/2}^2}\right)
  \simeq
  \frac{1}{32\pi}|U_{\gamma\nu}|^2
  \frac{m_{3/2}^3}{M_{pl}^2},  \label{eq:decay-width}
\end{eqnarray}
where $U_{\gamma\nu}$ represents the neutrino contamination into 
the ``photino''. This is approximately related to the neutrino mass as follows
\begin{equation}
   |U_{\gamma\nu}|^2 \simeq \cos^2 \theta_W \frac{m_{\nu}}{m_{\chi}},
  \label{eq:mixing}
\end{equation}
where $m_{\chi}$ is the mass of the bino-dominant lightest neutralino
and $\theta_W$ denotes the Weinberg angle. Using a representative
value $|U_{\gamma\nu}|^2 \simeq 7\times 10^{-13}$ which corresponds
to $m_{\chi} \simeq 80$ GeV, we find the gravitino
lifetime to be
\begin{eqnarray}
  \tau_{3/2}=\Gamma^{-1}(\tilde{G}\to \gamma \nu)
  \simeq 8.3 \times 10^{26} \sec \times
  \left(\frac{m_{3/2}}{1 \mbox{GeV}} \right)^{-3} 
   \left(\frac{|U_{\gamma\nu}|^2}{7\times 10^{-13}} \right)^{-1}.
\end{eqnarray}
Note that the lifetime becomes even longer as the gravitino mass decreases.
Thus we conclude that the gravitino is very long lived, whose lifetime is
much longer than the age of the Universe.

The long-lived gravitinos are generated in the early Universe. In the
standard big-bang cosmology, they were in thermal equilibrium and then
frozen out while they were relativistic. In this case their abundance
would be comparable to those of other light Standard Model particles.
If the Universe experiences inflationary expansion, then the primordial
abundance is completely diluted and the gravitinos are regenerated in
the thermal bath after the reheating. The abundance of the gravitinos
depends on the reheat temperature after the inflation
\cite{MMY}. Recently there has been claimed that a non-thermal
production mechanism during inflationary epoch may work to produce
more abundant gravitinos \cite{non-thermal}, 
though how it works will  depend very much on inflation models.
Thus we conclude that one can always consider a scenario of inflation
and subsequent reheating so that the gravitino abundance lies in a
right range where they constitute dark matter of the Universe.

Since the decay products contain photons, we have to next consider
constraints on the abundance of the photons produced.
The photon number flux induced by the gravitino decay has a peak at
the maximum photon energy $E_{\gamma}=m_{3/2}/2$ and the maximum flux there is
estimated as
\begin{eqnarray}
  F_{\gamma,\mathrm{max}}
  &=&E_{\gamma}\frac{dF}{dE_{\gamma}d\Omega}|_{E_{\gamma}=m_{3/2}/2}
  \nonumber\\
  &\simeq & \frac{n_{3/2,0}}{4\pi \tau_{3/2} H_0} 
\left(\frac{2 E_{\gamma}}{m_{3/2}} \right)^{3/2}|_{E_{\gamma}=m_{3/2}/2}
\nonumber \\
  &\simeq& 2.0\times 10^{-6}(\cm^2\cdot \,{\rm str}\cdot \sec)^{-1}
  \left(\frac{m_{3/2}}{1\GeV}\right)^2
  \left(\frac{\Omega_{3/2}}{0.3}\right)
  \left(\frac{h}{0.7}\right) 
  \left(\frac{|U_{\chi\nu}|^2}{7\times 10^{-13}} \right), \label{eq:flux} 
\end{eqnarray}
where $n_{3/2,0}=10.54({\rm cm}^{-3})(m_{3/2}/{\rm keV})^{-1}
(\Omega_{3/2}h^2)$ is the gravitino number density at present,
$\Omega_{3/2}$ stands for the gravitino mass density normalized by the
critical density of the Universe, and $H_0= 100 h$ km/sec/Mpc is the
Hubble constant.  We can obtain a constraint on the abundance
$\Omega_{3/2}$ by requiring that the flux
obtained in Eq.~(\ref{eq:flux}) does not exceed the observed diffuse 
photon background, which is fitted as \cite{EGRET}
\begin{equation}
F_{\gamma, obs}(E_{\gamma})
\simeq (1.5 \pm 0.3)
 \times 10^{-6} (\rm{cm^2} \cdot \rm{str} \cdot \rm{sec})^{-1}
\left(E_{\gamma}/ \rm{GeV} \right)
\end{equation}
for 20 MeV $<E_{\gamma} <$ 10 GeV.  We show our result in fig.1.  In
the $m_{3/2}$--$\Omega_{3/2}$ plane, the region excluded by the
consideration of the diffuse photon background is shown. We find that  only
a tiny region in the upper-right side is excluded. Here we conservatively 
took a 2 $\sigma$ error and imposed the constraint
\begin{equation}
  F_{\gamma,\mathrm{max}} \leq 2.1 
 \times 10^{-6} (\rm{cm^2} \cdot \rm{str} \cdot \rm{sec})^{-1}
\left(E_{\gamma}/ \rm{GeV} \right).
\end{equation}
The result implies that,
as far as the gravitino mass is less than about 1 GeV, $\Omega_{3/2}
\sim 0.1-1$ is allowed and thus the gravitino can constitute dark
matter. Note that actual limits on the gravitino mass depend on the
magnitude of the R-parity violating couplings, as we can see from
Eq.~(\ref{eq:flux}). 

Here we would like to briefly discuss another case where trilinear Yukawa
couplings are dominant sources of R-parity violation and neutrino masses.
In this case the neutrino masses are induced at one-loop level and  the 
gravtino decays to a neutrino and a photon also at the one-loop level. 
Therefore when we relate the decay width with the neutrino mass as we
did in Eqs.~(\ref{eq:decay-width}) and (\ref{eq:mixing}),
the decay width contains additional loop factor 
$\alpha/4\pi \sim 10^{-3}$ (with $\alpha$ the fine 
structure constant) compared to the previous case, and  the lifetime
will be longer by the inverse of the one loop factor.  Since the
photon flux coming from the gravitino decay is inversely proportional
to the lifetime, the constraint from the photon background becomes 
even weaker than the bilinear case.

In this paper, we have considered cosmology of the light
gravitino scenario when the R-parity is violated. Unlike the R-parity
conserving case, the decay of the NLSP becomes harmless because it
rapidly
decays to ordinary particles through the R-parity
violating interaction and its lifetime is much shorter than 1 sec.  On
the other hand, we showed that the lifetime of the gravitino LSP is
very long, typically by several orders of magnitude longer than the
age of the Universe. The decay products of the gravitino generically
contain photons, and thus the abundance may be constrained by the
observations of the diffuse photon background. Our analysis showed,
 when the bilinear R-parity violation induces the neutrino
mass which gives the neutrino oscillation solution of the atmospheric
neutrino anomaly, that  the long-lived gravitino can
constitute the dark matter of the Universe without conflicting the
observations of the diffuse photon background. As far as the gravitino
is heavier than $\sim 1$ keV, it behaves as a cold dark matter,
preferable from the arguments on the structure formation, while
for the mass of order 100 eV or less it becomes a warm dark matter which
is apparently ruled out \cite{PBMY}.  To summarize,  we
conclude that even when the R-parity is not conserved and thus the LSP
is unstable, it can be the dark matter, if it is the gravitino.

\acknowledgments
This work was supported in part by the Grant-in-Aid for Scientific 
Research from the Ministry of Education, Science, Sports, 
and Culture of Japan, 
on Priority Area 707 "Supersymmetry and Unified Theory of Elementary
Particles", and by the Grants-in-Aid No.11640246 and No.12047201.

\begin{figure}[t]
  \begin{center}
    ~\hfill
    \makebox[0pt]{
      \mbox{
        {
          \psfig{file=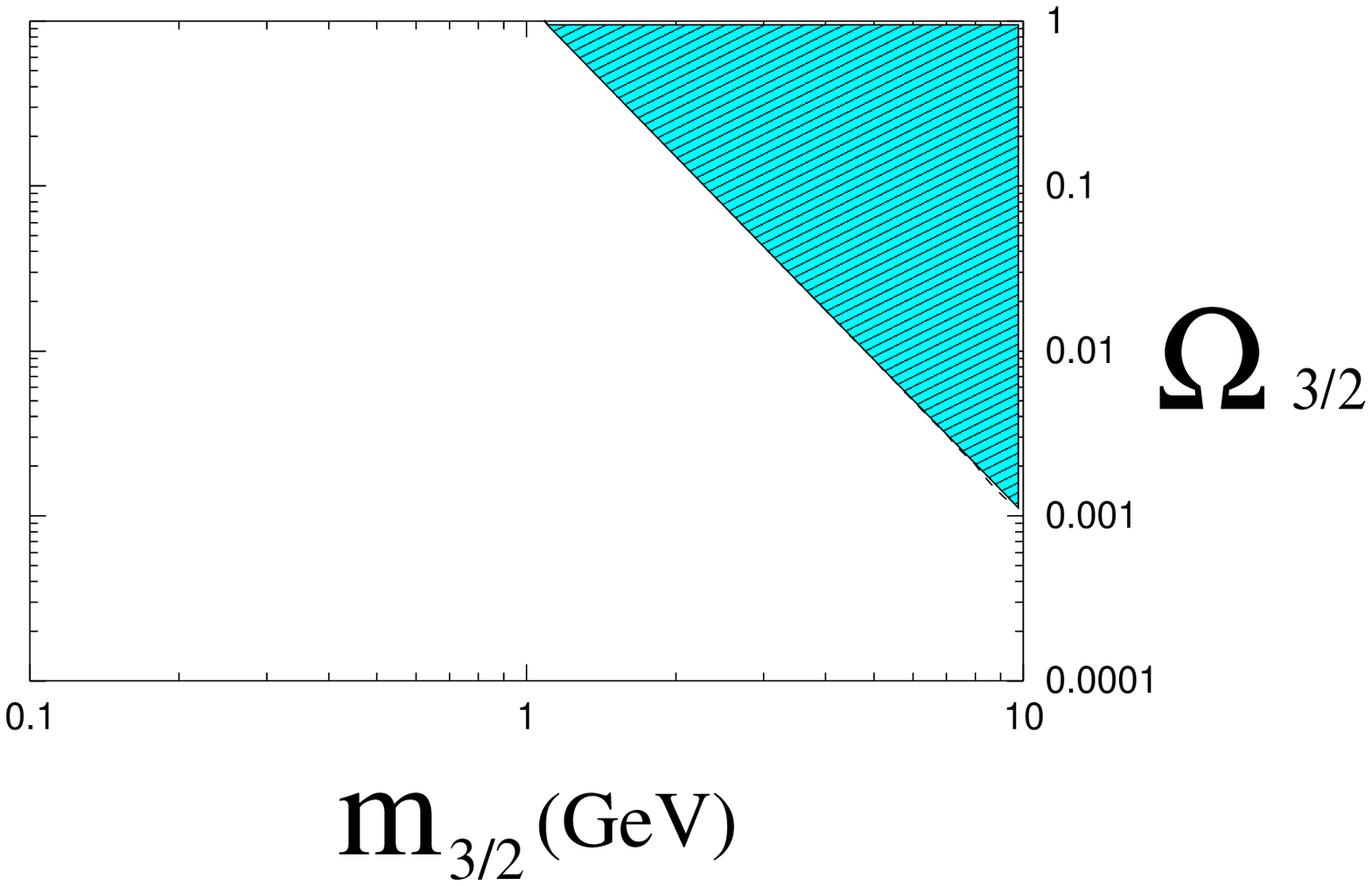,width=14cm}
          }
        }
      }
    \hfill~
  \end{center}
  \caption{Excluded region from the consideration of the diffuse
photon background in $m_{3/2}$--$\Omega_{3/2}$ plane. The shaded region
is excluded. We fix the neutrino-neutralino mixing $|U_{\gamma \nu}|^2 =
7 \times 10^{-13}$ and take 2 $\sigma$ error for the observed flux. }
  \label{??}
\end{figure}

\end{document}